# Observing 0D subwavelength-localized modes at ~100 THz protected by weak topology

Jinlong Lu[1]†, Konstantin G. Wirth[2]†, Wenlong Gao[1]†, Andreas Heßler[2], Basudeb Sain[1], Thomas Taubner[2,3]*, Thomas Zentgraf[1]*

[1] Paderborn University, Department of Physics, Warburger Str. 100, 33098 Paderborn, Germany.

[2] I. Institute of Physics (IA), RWTH Aachen University, 52074 Aachen, Germany.

[3] Jülich Aachen Research Alliance - Fundamentals of Future Information Technology (JARA-FIT), 52074 Aachen, Germany.

*Corresponding author. Email: taubner@physik.rwth-aachen.de; thomas.zentgraf@uni-paderborn.de

†These authors contributed equally.

**Abstract**

Topological photonic crystals (TPhCs) provide robust manipulation of light with built-in immunity to fabrication tolerances and disorder. Recently, it was shown that TPhCs based on weak topology with a dislocation inherit this robustness and further host topologically protected lower-dimensional localized modes. However, TPhCs with weak topology at optical frequencies have not been demonstrated so far. Here, we use scattering-type scanning near field optical microscopy to verify mid-bandgap zero-dimensional light localization close to 100 THz in a TPhC with nontrivial Zak phase and an edge dislocation. We show that due to the weak topology, differently extended dislocation centers induce similarly strong light localization. The experimental results are supported by full-field simulations. Along with the underlying fundamental physics, our results lay a foundation for the application of TPhCs based on weak topology in active topological nanophotonics, and nonlinear and quantum optic integrated devices due to their strong and robust light localization.

**Introduction**

Topological photonic crystals (TPhCs) provide robust light control against structural disorders and could relieve the crucial fabrication requirements limiting common photonic crystals (*1-4*). The idea originates from the seminal works by Raghu and Haldane, who generalized the topological band theory, discovered first in solid-state electron systems, to classical photonics based on the observation that band structure topology arising from waves in a periodic medium is regardless of the waves' classical or quantum nature (*5, 6*). Their discovery transferred the key feature of an electronic model - the Quantum Hall Effect - to the realm of photonics and they theoretically proposed the possibility of equipping classical waves with topological protection (*6*). With this extendedly robust protection of photons, lots of applications including robust light transmission (*7-10*), topological delay lines (*11*), robust lasing (*12, 13*), and quantum interference (*14-16*) can be achieved, promising further development of optical devices based on the topological design theory in the future.

On the other hand, state-of-the-art nanophotonics based on optical zero-dimensional (0D) localized modes with strong light-trapping is ubiquitously useful in applications such as nonlinear enhancement (*17*), photonic devices miniaturization (*18-21*), photonic or quantum chip integration (*19, 22*), and quantum electrodynamics (*23, 24*). However, the widely followed traditional strategy of using photonic crystal defect designs is notoriously fragile even to small structural changes, which lead to significant detuning of resonance frequency and mode volume (*25*). Hence, the implementation of robust topological protection of the 0D localized mode would boost the progress in these related fields. Recently, emerging seminal concepts with higher-order topological insulators (HOTIs) (*26-31*) and the Dirac vortex method (*13*) verified the promising routes for topological light localizations, e.g., showing several advantages in topological lasing. However, HOTIs need multiple domains of different unit cell designs to localize modes at the boundary between the domains. Another promising strategy to achieve topologically protected lower-dimensional localized modes lies in the design of weak topological insulators, which only exhibit these states for specific structural conditions (*32-36*). These weak topological insulators need only one, topologically nontrivial unit cell design in addition to, e.g., an intentionally introduced dislocation, whereas at the dislocation centers, strongly localized modes (0D) can be generated (*33, 34*). Still, the design and the experimental characterization of a weak topology protected localized mode has yet to be demonstrated at optical frequencies. First characterizations of TPhCs with a scattering-type scanning near field optical microscope (s-SNOM) have recently been reported for corner states in HOTIs (*28*) and edge states in a valley photonic crystal (*37*), while aperture-based near-field optical microscopy has been used to map edge modes (*9*). However, a spectroscopic study by s-SNOM of the evolution of the eigenmodes within a TPhC over a wide spectral range including the bandgap as well as bulk modes has yet to be demonstrated.

Here, we simulate and measure the topologically protected localized mode at optical frequencies in a TPhC design with weak topology as illustrated in Fig. 1. When the TPhC constructed by nontrivial unit cells possess a Zak phase vector along the same direction as the Burgers vector describing the edge dislocation, mid-bandgap 0D light localization is achieved, with a mode area ~ 0.17 $\lambda^2$ confined at the intentionally induced structural dislocation center. For differently extended dislocation centers, similar localization of light can be obtained, demonstrating the versatility and robustness of our design based on weak topology. For the experimental demonstration of the subwavelength localization, amorphous Silicon (a-Si) nanopillars are placed on a Chromium (Cr) film. The optical near-fields on top of the nanopillar arrays are measured using an s-SNOM at multiple frequencies around 100 THz. Employing a broadly tunable laser source (see Methods), we map the mode evolution within the TPhC over a range of about 35 THz, from the bulk states to the topologically protected mode within the bandgap. The measured electric field distributions show excellent agreement with the simulated results that are extracted from an eigenmode solver. Based on the 0D light localization at optical frequencies, our work proves the possibility to access and analyze such subwavelength interaction. Insights of the work open up a promising path for the application of weak topology and s-SNOM characterization in TPhCs at optical frequencies, enabling broad prospects in nonlinear optics, quantum optics, and photonic devices design and characterization.

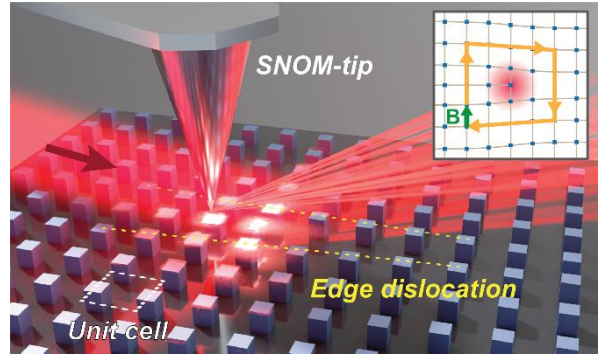

**Fig. 1. Schematic illustration of the s-SNOM measurement on the localized topological state.** With the edge dislocation formed by nontrivial unit cells, a strong field localization around the dislocation center is visualized by s-SNOM measuring directly the optical near-fields. Inset: Schematic of the dislocation point with corresponding Burgers vector **B**.

**Results**

To achieve 0D localization in the crystal dislocation design based on weak topology, we first calculate the band structure of the TPhC. In our design, we consider a square-lattice with a unit cell (constant period $P$ = 1334 nm) that consists of two rectangular pillars with a refractive index of 3.49 (corresponding to the value of a-Si determined from our measurement) placed on a perfect electric conductor as substrate (Fig. 2A). Using it as substrate reduces the requirements of fabricating dielectric nanostructures with a high aspect ratio.

The TPhC can be regarded as photonic realization of the 2D Su-Schrieffer-Heeger model, in which intracell and internal distances between two neighboring pillars are corresponding to the hopping amplitudes (*26, 38-41*). The topologically trivial ($d$ = 0 nm) and nontrivial ($d$ = 870 nm) unit cells here possess the same band structure with a full bandgap (94.4 THz — 101.6 THz) as shown in Fig. 2B. By considering the first bulk band, the topology of them is distinguished by a 2D Zak phase (*26, 39*):

$$\theta_i = \int dk_x dk_y \text{Tr}[\boldsymbol{A}_i], \quad i = \Gamma X \text{ or } \Gamma Y; \qquad (1)$$

where $\boldsymbol{A}_i = i\langle u_{\_k}|\partial_{k_i}|u_{\_k}\rangle$ is the Berry connection, $|u_{\_k}\rangle$ is the periodic Bloch function, and it is integrated over the first Brillouin zone. The Zak phase is a Berry phase of a system acquired by the periodic Bloch function going a closed path in the 1D Brillouin zone, and the 2D case here is a generalization to both directions ($k_x$ and $k_y$). The parity-time symmetry of the unit cell requires the Zak phase to be only 0 or $\pi$ (*42*). Equivalently, the electric field profiles' parity at

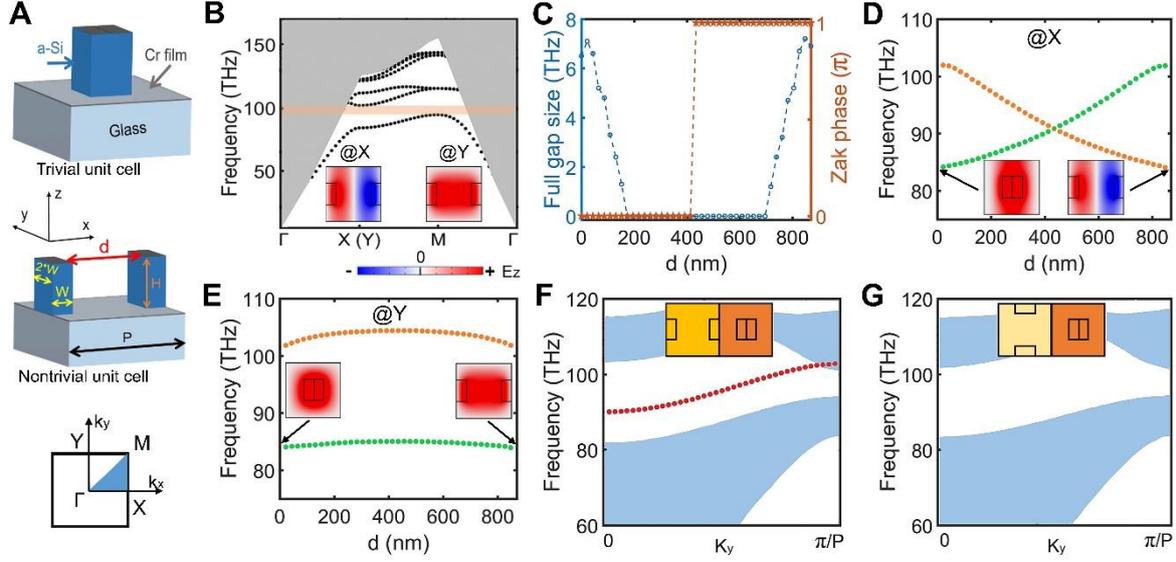

**Fig. 2. Band structure analysis of the TPhC. (A)** Schematic illustration of the topologically trivial unit cell ($d$ = 0 nm) and nontrivial unit cell ($d$ = 870 nm) made of a-Si nanopillars on a Cr film. **(B)** Corresponding band structure of the TPhC with $P$ = 1334 nm, $W$ = 232 nm and $H$ = 673 nm for both trivial/nontrivial unit cells (the size parameters are same in the whole text). The light cone is shaded in gray. **(C)** Evolution of the full bandgap size (shadowed with orange in **(B)**) and Zak phase of the lower band for different $d$ between the nanopillars. **(D and E)** Evolution of the two lowest bulk states at the X or Y point with different $d$. **(F and G)** Band structure of the supercells (with inset images showing the adjoining part) to verify the edge state, respectively. The red dotted line corresponds to the edge mode and the light blue shaded area to the bulk modes. The edge state is present with initial nontrivial/trivial unit cells **(F)**, while no edge state exists in the rotated nontrivial unit cell case **(G)**. Inset images in **(B, D and E)** show the $z$ component of electric fields ($E_z$) for the first band at different $k$ points.

the high symmetry points can indicate the Zak phase (*26, 27, 43*), by examining $\hat{M}_{x,y}|u_k\rangle = m|u_k\rangle$, in which $\hat{M}_{x,y}$ is the mirror operator, and $m$ is the mirror eigenvalue at mirror-symmetric Bloch moments. Since the fundamental bulk band at the Γ point is mirror-symmetric in the static limit, the Zak phase for $\theta_{\Gamma X}$ (equal to $\theta_{YM}$) is π only when $m$ = -1 at the X point (mirror antisymmetric), and the same applies for $\theta_{\Gamma Y}$ (equal to $\theta_{XM}$). Hence, the differences in the electric field distribution at the X and Y points shows that the nontrivial π Zak phase only emerges for $\theta_{\Gamma X}$ (inset image of Fig. 2B). The result is identical to the numerical calculations by the Wilson loop approach in which complex electromagnetic fields obtained from simulations, substitute the eigenstates in Equation (1) (details in Section 1 of the supporting information). By dividing the closed-loop in the Brillouin zone into a larger number of small segments, the Wilson-loop approach provides a gauge invariant strategy to calculate topological invariants from numerical simulation, and the Zak phase is the summation of the Berry phase in all of the small segments. The full bandgap size together with the topological phase transition through changing $d$ is further illustrated in Figs. 2C, D, E, with $d$ = 435 nm as

the topological transition point, where the bandgap closes and reopens. Note that the parity inversion only occurs at the X point and is not observed at the Y point.

We check the bulk-edge correspondence — which is used to determine the presence of edge states at an abrupt interface — as shown in Figs. 2F, G by studying a single supercell with periodic boundary conditions only in y-direction. The supercell is formed by arranging seven (rotated) nontrivial unit cells followed by seven trivial unit cells along the x-direction (the inset images show the adjoining section). Here, we find a unidirectional transport mode within the bandgap when the a-Si nanopillars in the trivial and not-rotated nontrivial unit cells are arranged as shown in the inset in Fig. 2F, corresponding to the nontrivial Zak phase $\theta_{\Gamma X} = \pi$. In contrast, no edge state is present if we rotate the nontrivial unit cell by 90 degrees, as the Zak phase $\theta_{\Gamma Y} = 0$ in such a design.

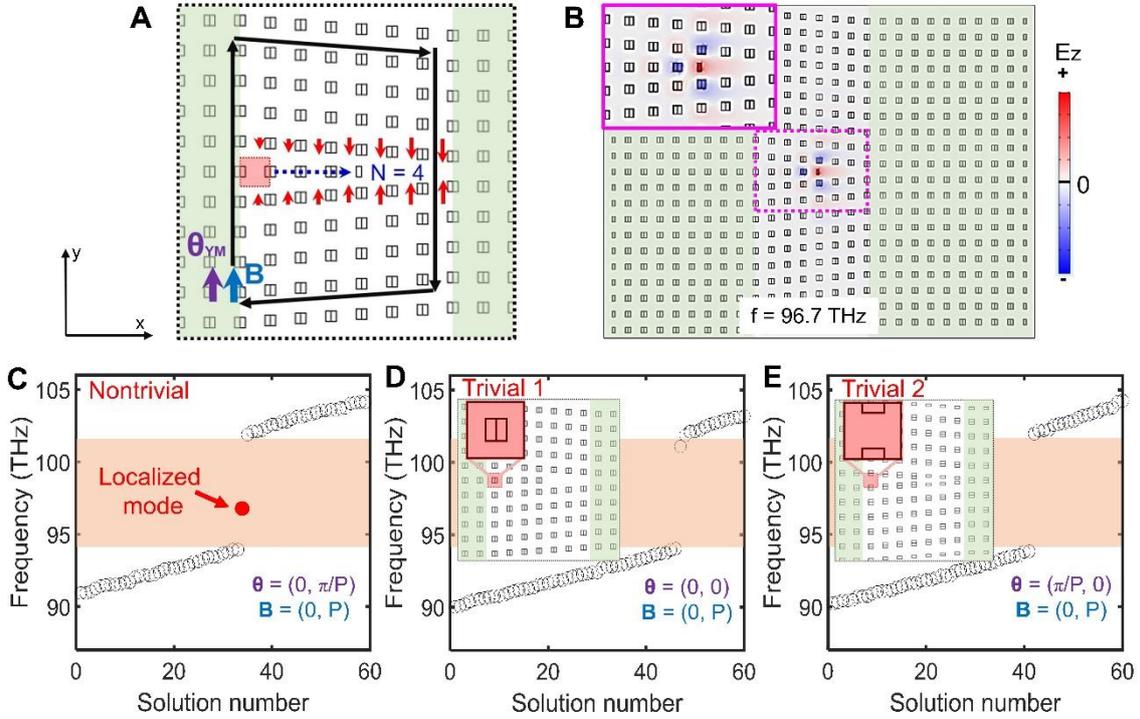

**Fig. 3. Comparison of three kinds of TPhCs.** (**A**) Illustration of the defect formation with the nontrivial design as an example, with $\theta = (0, \pi/P)$ and $\mathbf{B} = (0, P)$ shown by the two solid arrows. The uniform area (corresponding to the uniformly arranged red shadowed unit cell in each design) is shadowed with green. The dark blue dashed arrow indicates the extension of the dislocation center with $N = 4$ more unit cells. (**B**) $E_z$ field distribution obtained from the eigenmode calculation of the marked in-gap localized mode in the spectrum of the *Nontrivial* design. The dotted purple line marks the area for the magnified view shown as inset. (**C**) Spectrum of the *Nontrivial* design with the localized mode in the bandgap. (**D**) Layout at the dislocation center and spectrum of *Trivial 1* design with $\theta = (0, 0)$. (**E**) Layout at the dislocation center and spectrum of *Trivial 2* design with $\theta = (\pi/P, 0)$. The unit cell in the red box (**A**, **D** and **E**) shows the starting position of the dislocation center extension for each design. The bandgap region obtained from the band structure of the unit cells is shadowed in orange.

Next, we introduce a structural defect in a TPhC consisting of topologically nontrivial unit cells (with 26 · 21 uniformly arranged unit cells) as shown in Fig. 3A: First, we partly remove the centerline of the unit cells (from the one shadowed in red, which sits at the 11$^{th}$ line in y-direction, and the 10$^{th}$ unit cell in x-direction) in the uniform part of the TPhC. Then, all the units above and below the removed section are shifted linearly along the y-direction towards the centerline (see red arrows) to compensate for the removal. This procedure creates an edge dislocation with a single pillar in the middle as the dislocation center. A closed-loop, as shown by black arrows, results in a Burgers vector **B** = (0, *P*) (the light blue arrow) along the y-direction in real space, which can be used to describe this dislocation. Finally, the dislocation center is further extended with *N* more unit cells (here *N* = 4, the number of unit cell from the one shadowed in red) as shown in the direction of the dashed arrow to reduce the mode scattering around the pillar.

Here, the weak topological invariant *Q* (which equals the number of localized modes) contains both parameters from the band structure topology and crystal dislocation (*33*):

$$Q = \frac{1}{\pi}\boldsymbol{\theta} \cdot \mathbf{B} : \mod 2 \qquad (2)$$

where $\boldsymbol{\theta} = \left(\frac{\theta_{XM}}{P}, \frac{\theta_{YM}}{P}\right)$ represent the Zak phase information of the band below the gap along the XM and YM line, respectively, in the Brillouin zone; **B** describes the crystal edge dislocation in real space. *Q* can only be 0 (no localized mode) or 1 (with a localized mode). According to this weak topological invariant *Q*, the case when the nonzero Zak phase vector $\boldsymbol{\theta}$ and the Burgers vector **B** are parallel to each other can result in the topologically protected localized mode. This topological concept was termed "weak" to distinguish it from the well-known strong topological insulators (where "strong" is always omitted) which are solely characterized by a quantized scalar topological invariant (*32, 34*).

To verify that the localization only appears in the nontrivial design based on weak topology, we study the spectra and electric field profiles of three kinds of TPhCs. Based on the 2D Zak phase or $E_z$ field distribution in Fig. 2B (with the field distribution at the M point the same as at the X point, not shown), the nontrivial unit cell gives rise to $\theta_{YM} = \pi$, while $\theta_{XM} = 0$ as for the parity inversion difference; and both $\theta_{YM}$ and $\theta_{XM}$ are equal to 0 for the trivial unit cell. This results in the three TPhCs: *Nontrivial* (Figs. 3A-B), constructed by the initial nontrivial unit cell with $\boldsymbol{\theta}$ parallel to **B**, and $\boldsymbol{\theta} = (0, \pi/P)$; *Trivial 1* (Fig. 3D), constructed by the trivial unit cell with $\boldsymbol{\theta} = (0, 0)$; *Trivial 2* (Fig. 3E), constructed by the 90-degree rotated nontrivial unit cell with $\boldsymbol{\theta} = (\pi/P, 0)$ perpendicular to **B**. Note here, the design strategy for these two trivial TPhCs is similar to that of the nontrivial design, explained above regarding Fig. 3A.    Coinciding with our previous assumptions, a localized mode (*f* = 96.7 THz) in the bandgap (94.4 THz — 101.6 THz) around the dislocation center is only found for the nontrivial design as shown in Figs. 3B-C. In contrast, for the two comparative trivial designs (*Trivial 1* and *Trivial 2*), no localized state inside the bandgap is found in the spectra (Figs. 3D-E), verifying our design strategy to realize the in-bandgap localized mode with the nontrivial Zak phase and edge dislocation. The additional requirement of a certain lattice arrangement in addition to the nontrivial Zak phase indicates that the localization is indeed protected by weak topology.

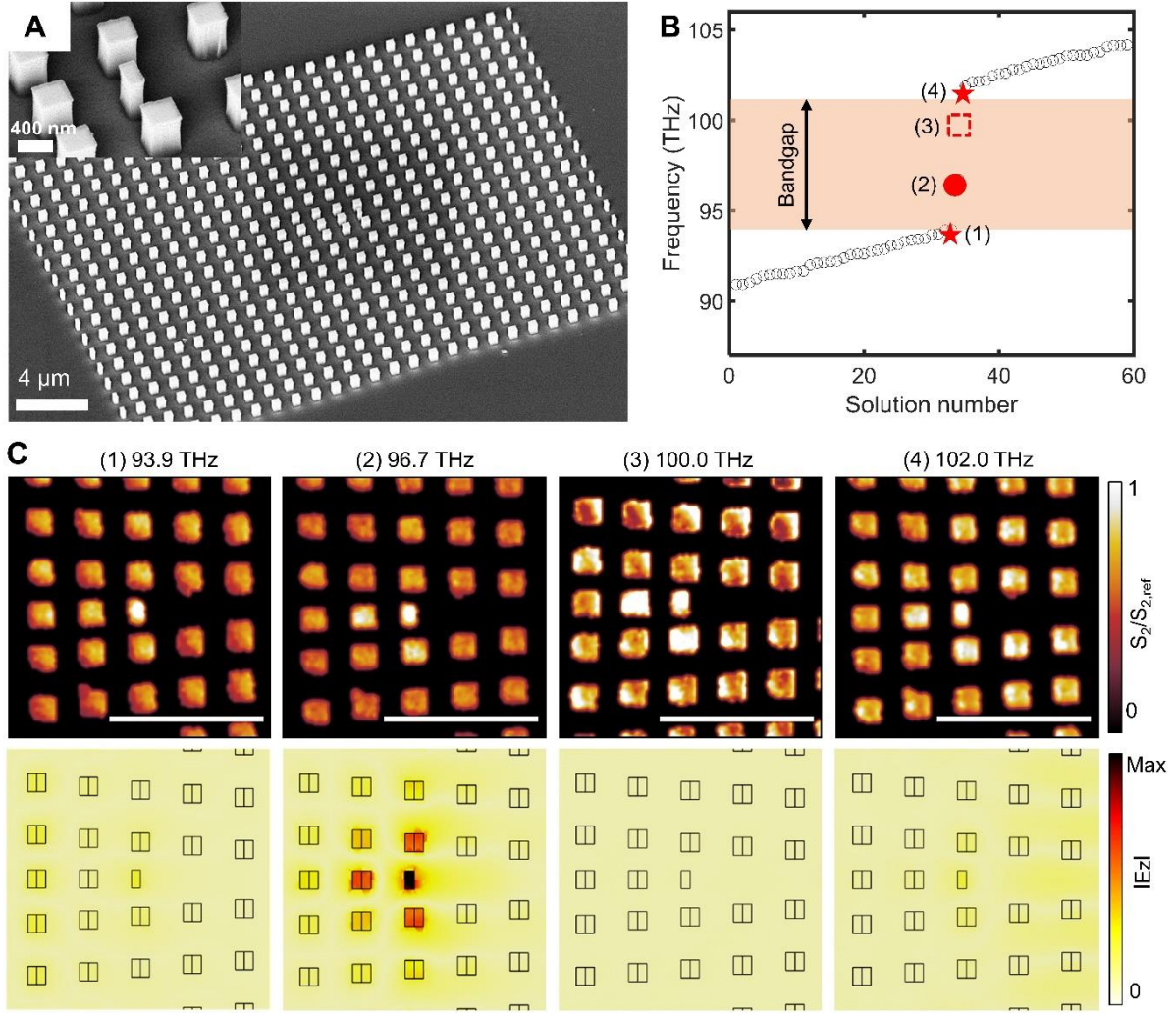

**Fig. 4. Frequency-dependent near-field results.** (**A**) Scanning electron microscopy image of the nontrivial TPhC (inset: magnified view of the dislocation center). (**B**) Corresponding spectrum with the measured frequencies marked with red markers. (**C**) Normalized scattering-type scanning near-field optical microscopy (s-SNOM) amplitude images to show the measured electric field distribution (top) and the simulated $E_z$ fields (absolute value) obtained from the eigenmode solver (bottom) for each frequency. The subwavelength localization with strong fields around the single pillar at the dislocation center is observed in the bandgap at $f$ = 96.7 THz, while the fields for other frequencies show weaker localization and broader distributions. Scale bars are 4 μm.

The electric field profile of the designed localized mode in Fig. 3B suggests a mode area of ~ 0.25 $\lambda^2$ ($\lambda$ is the wavelength in vacuum) around the dislocation, which indicates a more strongly confined mode compared to other topological designs like HOTIs and the Dirac vortex method (*13, 26, 28*). Here, the mode area of the localization is calculated with the formula:

$$A_{eff} = \frac{\left(\int |E_z(r)|^2 d^2 r\right)^2}{\int |E_z(r)|^4 d^2 r}, \tag{3}$$

where $|E_z(\mathbf{r})|$ is the $z$ component of the electric field, and the integration is performed on top of the structure (in the plane $z$ = 680 nm).

To realize the designed TPhC above, we fabricated a-Si pillars on a glass substrate coated with a 100 nm-thick Cr film as shown in the SEM image in Fig. 4A. The target unit cell dimension corresponds to the simulated size (lattice constant $P$ = 1334 nm, height $H$ = 673 nm, size of each single pillar $W \cdot 2W$ = 232 nm $\cdot$ 464 nm). In Figure 4B of the related spectrum, the four investigated frequencies are indicated by the red markers. $E_z$ fields (in the plane at z = 680 nm) from the eigenmode solver corresponding to the near-field distribution at each frequency are shown below the experimental s-SNOM results in Fig. 4C (for explanation see methods). A strongly confined field localized around the single pillar at the dislocation center is measured for the frequency of the state in the bandgap ($f$ = 96.7 THz), whereas the other frequencies show a weaker localization and broader field distributions. The simulations reproduce the measurement results well (further comparison of electric fields can be found in Section 3 of the supporting information). Interestingly, we also observe the two eigenmodes at the band edge ($f$ = 93.9 THz and $f$ = 102.0 THz), where a slight localization around the center pillar can be seen. At 100 THz, where no eigenstate is expected within the bandgap, the measured signal on the pillars is comparable to the center pillar. The light is scattered non-resonantly by the array. That the pillars at the center appear slightly uniformly brighter can be explained by the spectral resolution of ~ 1.5 THz of the tunable laser source. Therefore, the eigenstates within this spectral range around 100 THz, especially at the band edge ($f$ = 102.0 THz) and in the band gap ($f$ = 96.7 THz) can also be slightly excited. This overlap is reflected in the measured signals, like the middle pillar in Fig. 4C, giving rise to bright regions in the measurements even though the actual field values might be relatively small (see simulations).

The spatial localization of the mid-bandgap eigenmode can be visualized by simulating the near-field distribution of the TPhC with excitation by an obliquely incident plane wave. The calculations show a near-field enhancement of more than 30 times above the pillars ($z$ = 680 nm), and the electric field distribution agrees well with the s-SNOM measurements (for more details see supporting information Section 3). Based on these results, the designed mid-bandgap localization and bulk states at optical frequencies could be demonstrated and verified by s-SNOM measurements, making it a promising tool to characterize TPhCs at optical frequencies (more details in Section 3 of the supporting information) (*28*).

To further demonstrate the versatility of our weak topology design to achieve the 0D localized state, we consider several design layouts based on the nontrivial unit cell, but with a different number $N$ of unit cells extending the dislocation center (see illustration of $N$ in Fig. 3A, $N$ = 3, 5, 6 for the three layouts here) as shown in Fig. 5. The calculated spectra (left column) still show a localized mode inside the bandgap as marked by the red dot, and similar field confinement around the dislocation center can be observed (center column). The smallest mode area of ~ 0.17 $\lambda^2$ (see Equation (3)) is achieved for the design with $N$ = 6 due to the scattering suppression with a smaller vacant area in the array (largest mode area ~ 0.36 $\lambda^2$ for $N$ = 3). To verify the simulation and to show the variation of the mode localization in the different layouts, the arrays were fabricated as before and measured with s-SNOM at the marked frequencies of the localized eigenstate inside the bandgap. As shown in the normalized field distributions (right column in Fig. 5), a strong decrease in the relative field amplitude on the surrounding pillars with respect to the single pillar can be observed with increasing $N$. The spectra and localized field distributions together show the prominent advantages of the weak topology design for tuning of the 0D localization properties, i.e., frequency, relative localization position, and mode area.

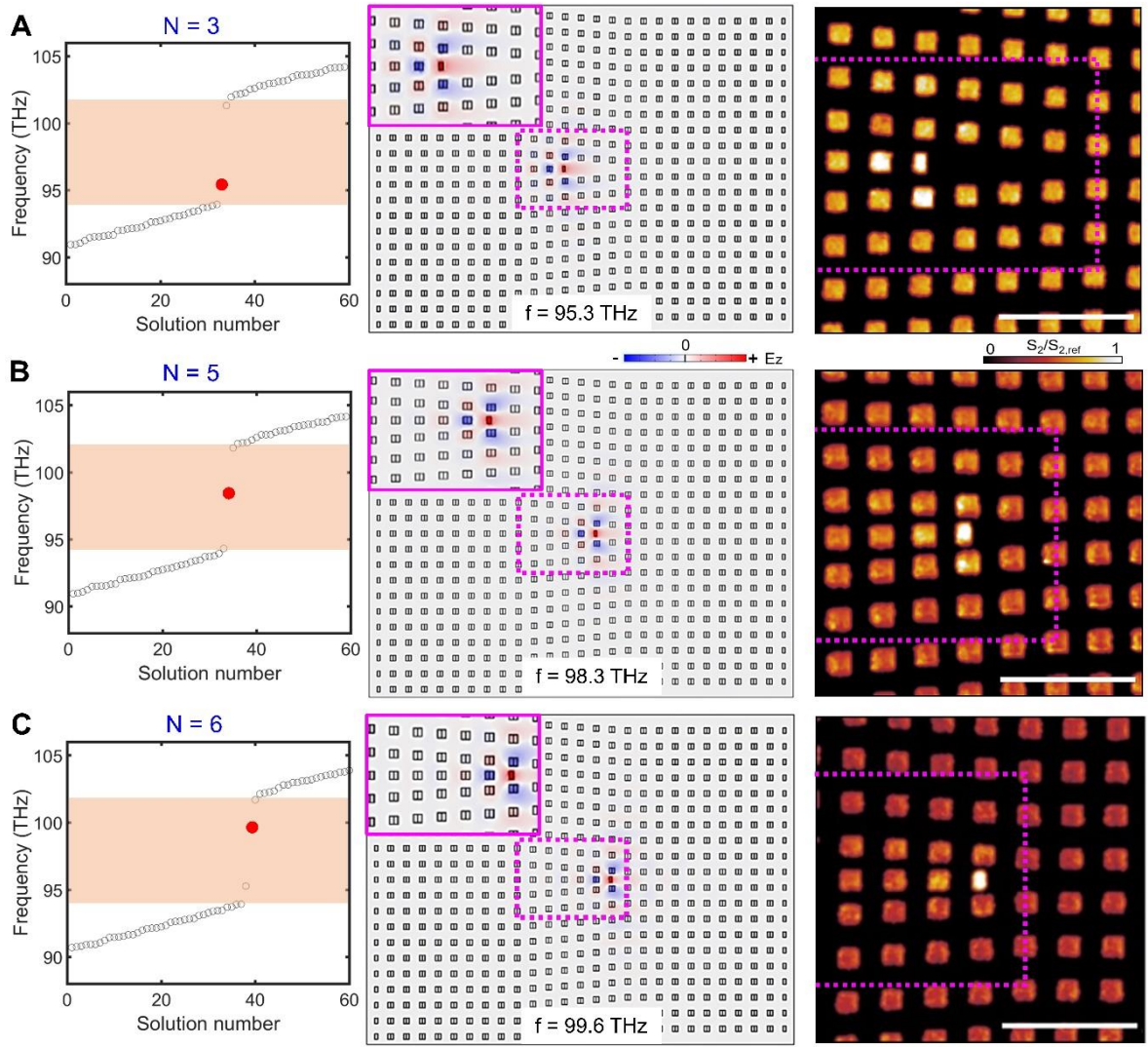

**Fig. 5. Versatility of the weak topological design.** Demonstration of the versatility of the design to achieve the 0D localized mode with three different unit cell extensions at the dislocation center for (**A**) $N = 3$, (**B**) $N = 5$, and (**C**) $N = 6$, respectively. The left column shows the spectra, middle column shows the simulated field distribution from the eigenmode solver, right column shows the normalized s-SNOM field distribution. Simulated and experimental field distributions show similar localization at the mid-bandgap (red dot in each spectrum) around the dislocation center. The localization becomes stronger with larger $N$. The dotted lines mark the area of the magnified inset in the simulation images, corresponding to the section shown in the measurements. Scale bars are 4 μm.

In stark contrast to the nontrivial designs, no topologically localized mode can be found in all the trivial TPhCs as shown in the supporting information (Section 4). As only topological

Page **9** of **24**

modes can appear and follow the topological phase transition. The spectra and field distributions comparison between the corresponding trivial and nontrivial TPhCs indicates the modes that only exist in the nontrivial design are topological modes. The versatility of the weak topological design provides a flexible strategy to access and further adjust the localization of light fields. Together with the robust nature of topological protection (see robustness of the mode in supporting information Section 5), it is highly promising for future photonic device design.

**Discussion**

The weak topology applied in our work enables a mid-bandgap 0D localized mode, which is present when the vector describing the nontrivial Zak phase is parallel to the Burgers vector of the edge dislocation according to the defined weak topological invariant. Furthermore, the designed optical localization with a subwavelength confined area is experimentally verified by s-SNOM measurements of a TPhC made of a-Si nanopillars placed on a layer of Cr. Based on the versatility of the weak topology, similar in-bandgap 0D localizations with a confined field at the intentionally induced structural dislocation center are designed with different dislocation extensions. With a larger number of unit cell extensions at the dislocation center, a reduced mode area is observed in the measurements.

Confining light at optical frequencies to a point (0D) is promising for applications in nonlinear optics, quantum optics, and miniature active photonic device integration. Our study demonstrates a feasible strategy for accessing and tuning a topologically protected 0D localized state at infrared frequencies. We also show that s-SNOM is a valuable tool for characterizing TPhCs with nanoscale resolution at optical frequencies. Going further, programmable phase-change materials (*44-47*) with tunable refractive index could be included into TPhCs to realize robust and active topological nanophotonics. Furthermore, the rapid progress of nanotechnology makes it possible to equip the defect center with quantum emitters, nanocrystals or molecules, which is highly promising for novel hybrid nanophotonic devices for single photon generation.

**Methods**

*TPhC fabrication:* First, a thin Cr film with a thickness of 100 nm was deposited on a cleaned glass substrate by electron beam evaporation. An amorphous silicon (a-Si) layer of thickness 673 nm was deposited by plasma-enhanced chemical vapor deposition (PECVD), immediately on top of the Cr film. The sample with deposited a-Si on top of the Cr film was then used to perform the subsequent processes to transfer the desired patterns onto a-Si. A poly-methyl-methacrylate (PMMA) resist layer was spin-coated onto the a-Si film and baked on a hot plate at 170 °C for 2 mins to remove the solvent. The desired structures were patterned by using a standard electron beam lithography and developed in 1:3 methyl isobutyl ketone (MIBK): isopropyl alcohol (IPA) solution. Next, another 20 nm thick Chromium (Cr) mask was deposited by electron beam evaporation. After a lift-off process in hot acetone, the patterns were transferred from PMMA to Cr. Finally, the structures were transferred onto a-Si using an inductively coupled plasma reactive ion etching (ICP-RIE) and the following removal of the Cr mask by a commercially purchased Cr-etch solution.

*Simulation:* The band structure and edge state were calculated by using a commercially available COMSOL Multiphysics software with the RF module using the eigenmode solver. The perfect electric conductor boundary condition was used for the bottom and the periodic boundary condition was used for each pair of the unit cells or supercells. The spectra and corresponding near-field distributions were simulated using the eigenmode solver in the same way, but all with scattering boundary conditions outside the structure. For the refractive index of the a-Si, we use 3.49 obtained from ellipsometric measured data in the range of 600-1500 nm fitted with a Lorentz model.

*Near field measurement:* Scattering-type scanning near-field optical microscopy (s-SNOM) records the scattering amplitude and phase of the optical near-fields at the AFM tip a certain laser frequency by lock-in detection (*48*). The signal is then demodulated at higher harmonics of the tip's oscillation frequency.

Here, we deployed an s-SNOM set up by Neaspec GmbH in pseudo-heterodyne detection mode in combination with an $LN_2$ cooled InSb detector optimized for the wavelength range up to 5.4 μm (Infrared Associates) to record amplitude and phase data simultaneously. The s-SNOM AFM was operated in tapping mode at oscillation frequencies between 220 and 270 kHz, with commercially available metal-coated AFM probes. The tapping amplitude is 40-60 nm. The laser source is a commercially available OPO/OPA pulsed laser (Alpha Module by SI-Instruments) tunable between 65 to 217 THz. The laser system has a repetition rate of 42 MHz and a pulse duration of up to 1 ps. This provides a spectral resolution of 1.5 THz and allows for recording images in s-SNOM in pseudo-heterodyne detection mode around 100 THz (~ 3 μm) and above as recently shown (*49*). We extracted the signal from the second demodulation order amplitude. The spectral resolution of the laser leads to a smearing of the states compared to the simulations which work at a single frequency. Thus, not every eigenstate is perfectly localized as expected from simulations.

*Evaluation of s-SNOM measurements:* Between the pillars, the AFM probe picks up the strong background signal of the Cr substrate, thus the substrate area is not used for evaluation and set to zero. Instead, the second-order normalized amplitude images in Fig. 4 and 5 were obtained by evaluating the amplitude signal on top of the pillars. The top pillar position was determined by setting a threshold value in AFM images and overlayed with the corresponding optical images, setting the surrounding to zero. Afterward, the signal was averaged over two neighboring pixels with a gaussian filter to smooth the pixels and normalized to the signal on top of the middle pillar (dislocation center with a single pillar) to highlight the field distribution. The step-by-step data analysis is shown in supporting information Section 6.

**Acknowledgments**

**Funding:** This project has received funding from the European Research Council (ERC) under the European Union's Horizon 2020 research and innovation program (grant agreement No. 724306) and the Deutsche Forschungsgemeinschaft (DFG No. ZE953/11-1, DFG No. TA848/7-1, SFB 917 "Nanoswitches").

**Author contributions:** W.G. conceived the idea. J.L. and W.G. performed the numerical simulation and calculations. J.L. and B.S fabricated the sample. K.W. and A.H. did the near field measurement and data analysis. T.T. and T.Z. guided the research. All authors contributed to the discussion of the results and the preparation of the manuscript.

**Competing interests:** The authors declare that they have no competing interests.

**Data and materials availability:** All data needed to evaluate the conclusions in the paper are present in the paper and/or the Supplementary Materials.

# Supplementary Information

## 1. Zak phase calculation with Wilson loop approach

The topology of a band in a photonic system can be defined by the topological invariant linked with the Berry phase, which is the geometric phase acquired by the periodic Bloch function going a closed path in Brillouin Zone (BZ) *(1)*. The Berry phase is defined as:

$$\gamma = \oint d\mathbf{k} \cdot A(\mathbf{k}) \qquad \text{S(1)}$$

where $A(\mathbf{k}) = i\langle u_{\_k} | \nabla_k | u_{\_k} \rangle$ is the Berry connection, $|u_{\_k}\rangle$ is the periodic Bloch function and the integration in a closed path in BZ.

Direct calculation of the Berry phase meets the problem of gauge uncertainty and is inconsistent with numerical calculation as the continuous evolution of the Bloch functions in the BZ are required (but the continuous evolution is impractical in numerical simulation). This calculation problem can be solved with the well-known Wilson-loop approach *(2-3)*. The Zak phase is a Berry phase of a specifically chosen 1D BZ (more specifically, a closed-loop for the given $k_x$ or $k_y$) *(4)*. Here, we describe how the Zak phase that is used in the main text can be calculated with the Wilson loop approach, a similar method can be used for the calculation of other topological invariants *(2-3, 5)*.

We rewrite the Zak phase as the following:

$$\theta = \oint d\mathbf{k} A_{n,k} \qquad \text{S(2)}$$

using the Berry phase of S(1), $\mathbf{A}_{n,k} = i\langle u_{\_k} | \nabla_k | u_{\_k} \rangle$ is the Berry connection for the n$^{th}$ band, $|u_{\_k}\rangle$ is the periodic Bloch function, and the integration is performed over a closed-loop for that specifically chosen $k_x$ (with $-\pi \leq k_y < \pi$ as the loop for calculation) or $k_y$ (with $-\pi \leq k_x < \pi$ as the loop) in BZ.

By dividing the closed-loop of 1D BZ at the specific $k$ point (the point corresponding to the direction of the Zak phase that we want to calculate) into $N$ (large enough) small segments, we can approximate the integral in S(2) as the summation of the contributions from each small segment. In a small segment $j$ from $k_j$ to $k_{j+1}$ (range of $j$ is from 1 to $N$), we have $e^{-i\theta_j} \approx 1 - i\theta_j = 1 - i\mathbf{A}_{n,k}\delta\mathbf{k}$ if $N$ is sufficiently large, and the differentiation of the Bloch functions over $k$ is replaced by the finite differences. Then the Berry phase $\theta_j$ in segment $j$ is given by $e^{-i\theta_j} = \langle u_{n,\mathbf{k}_j} | u_{n,\mathbf{k}_{j+1}} \rangle$.

The Zak phase corresponding to this $k$ point is then the summation of the Berry phase in all small segments:

$$e^{-i\theta} = \prod_{j=1}^{N} e^{-i\theta_j} = \prod_{j=1}^{N} \langle u_{n,\mathbf{k}_j} | u_{n,\mathbf{k}_{j+1}} \rangle. \qquad \text{S(3)}$$

This method does not require the continuous evolution of the Bloch functions in the BZ, as each of them appears twice in the above product. The gauge-invariant calculation here is the basic starting point of the Wilson-loop approach and allows for the topological invariant calculation from numerical simulation.

In real calculation with classical waves from numerical simulation, if no magneto-electric coupling is considered, the Berry connection of the electromagnetic wave in an artificial structure is given by $A^E_{n,k} = i\langle u^E_n(k)|\varepsilon(r)|\partial_k u^E_n(k)\rangle$, which is the contribution from the electric fields for the $n^{th}$ band. Then, with $u^E_n(k)$ for each $k$ point obtained from numerical simulations in COMSOL Multiphysics, the Zak phase is the imaginary part of the logarithmic of all the multiplication.

More specifically, the 2D Zak phase for $\theta_{\Gamma X}$ in the main text is calculated for $k_y = 0$, while the loop is $-\pi/P \leq k_x < \pi/P$; $\theta_{\Gamma Y}$ is calculated for $k_x = 0$, while the loop is $-\pi/P \leq k_y < \pi/P$. The results give $\theta_{\Gamma X} = \pi$, $\theta_{\Gamma Y} = 0$. $\theta_{XM}$ is calculated for $k_x = \pi/P$, while the loop is $-\pi/P \leq k_y < \pi/P$; $\theta_{YM}$ is calculated for $k_y = \pi/P$, while the loop is $-\pi/P \leq k_x < \pi/P$. The results give $\theta_{XM} = 0$, $\theta_{YM} = \pi$.

## 2. Determination of the field enhancement for the nontrivial TPhC

The localized state (design for $N = 4$) from the eigenmode solver can be visualized in COMSOL with the background field excitation, and the field enhancement is determined from simulation. The background electric field direction and polarization were set according to the measurement condition (55° with respect to surface normal) with a linear polarized plane wave. A hemispheric perfectly matched layer (PML) as the boundary of the simulation domain. A perfect magnetic conductor (PMC) was used for the x-z plane (perpendicular to the bottom) here due to the mirror symmetry of the layout. From the electric field above the pillars ($z = 680$ nm) at the localized frequency of $f = 96.7$ THz, a field enhancement of 30 times can be found around the pillar at the dislocation center as shown in Fig. S1. The field pattern around the pillar at the dislocation center looks similar to the results from the eigenmode solver. We also checked the $E_z$ field distribution for the other frequency at $f = 93.9$ THz: A rather weak field distribution over a broad area is observed, corresponding to the bulk modes based on the simulation.

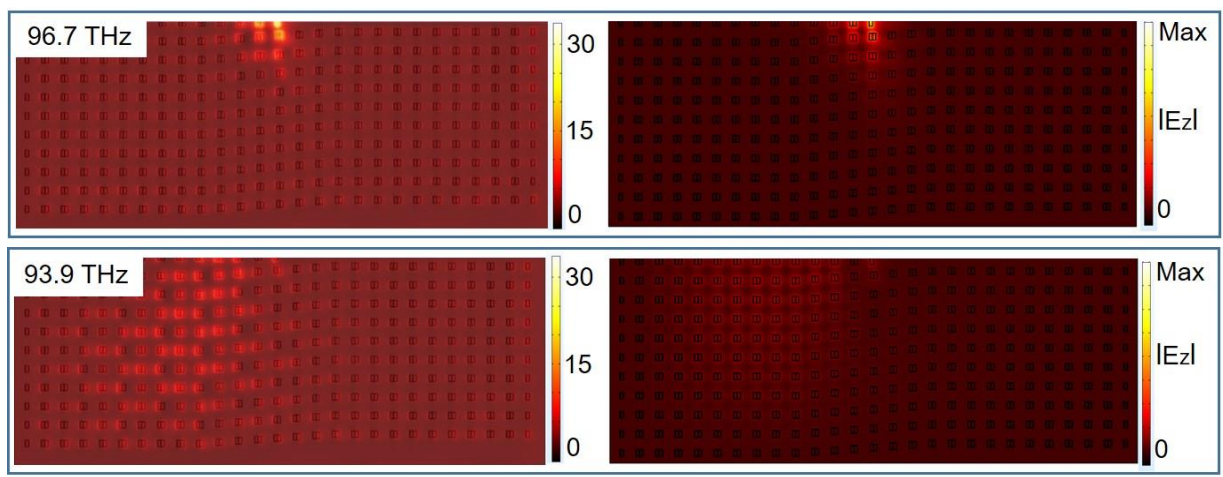

**Fig. S1.** Extracted electric field distribution from the simulation at $z = 680$ nm with background excitation (left column) and results from the eigenmode solver (right column). Both simulated results show similar electric field distributions.

## 3. Additional measurements and simulation results for the nontrivial TPhC

In Fig. S2, normalized second demodulation order ($S_2$) amplitude images for more measured frequencies and an additional position are shown (the same sample as in Fig. 4 in the main text). $f$ = 83.3 THz and 119.9 THz correspond to frequencies far away from the band gap. The results around the dislocation center show similar field distributions over the whole area and indicate bulk modes as shown below. The image for $f$ = 96.7 THz (localized mode) shows a region far off the dislocation (outside the red box in (c)), showing homogenous field distribution on the pillars, indicating the absence of enhanced fields at these positions off the dislocation center like the simulation.

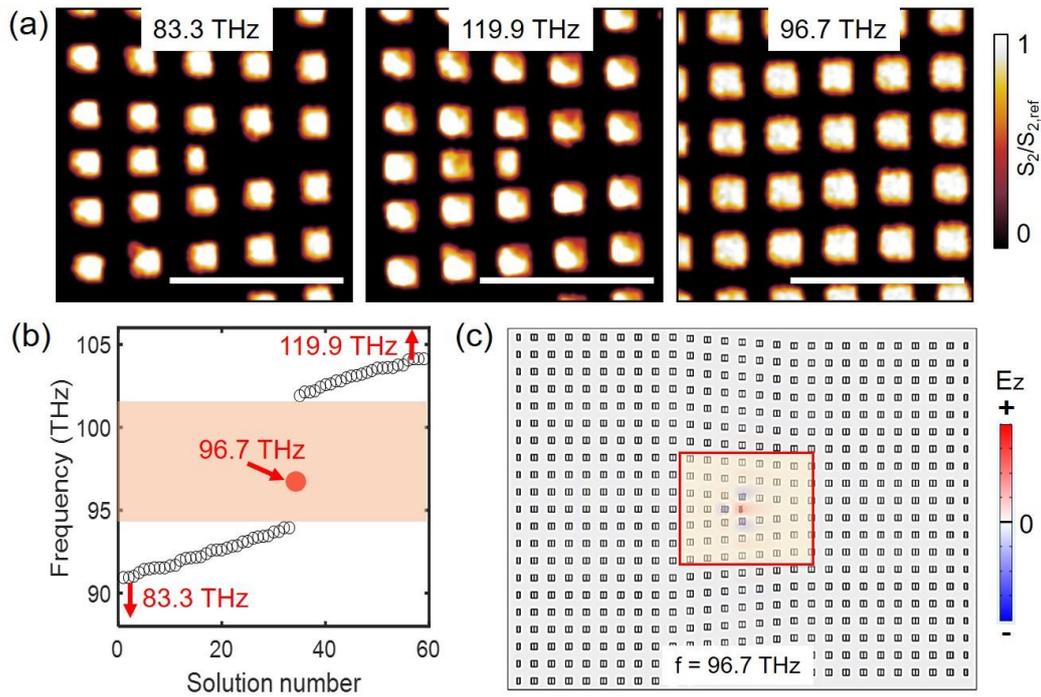

**Fig. S2.** (a) Normalized $S_2$ amplitude images from measured results for more frequencies or position ($N$ = 4). (b) Illustration of bulk modes for the two frequencies far away from the bandgap. (c) Illustration of the measured area off the dislocation center for the measured localized frequency at $f$ = 96.7 THz. The fields on the pillars are homogenous for all three images, indicating the absence of enhanced fields. Scale bars are 4 μm.

In Fig. S3, we show the $E_z$ field distribution from the eigenmode solver with the $E_z$ field normalized to the maximal value in each field, which is a similar way to the data normalization from s-SNOM results (the electric field is normalized to the maximal value from each frequency). All the field distributions for each frequency agree very well with the measurement (Fig. 4 in main text).

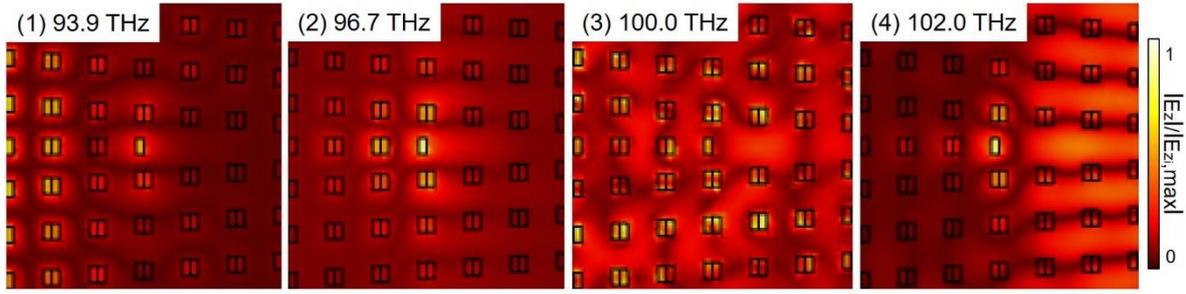

**Fig. S3.** Simulated $E_z$ fields from the eigenmode solver for each frequency (top row). The field distribution is shown with the electric field normalized to the maximal value in each frequency, similar field distributions are observed for the measured results in main text (Fig. 4).

Since an s-SNOM set-up with pseudo-heterodyne detection was, amplitude and phase were recorded simultaneously. The images with phase information were corrected for a linear phase drift. A drift in the 100 THz phase image can still be observed, where the phase drift was apparently non-linear (Fig. S4). The background was set to zero, like the procedure for the amplitude images presented in Section 6 of the Supporting Information. The phase signal does not show a pronounced difference between the different frequencies.

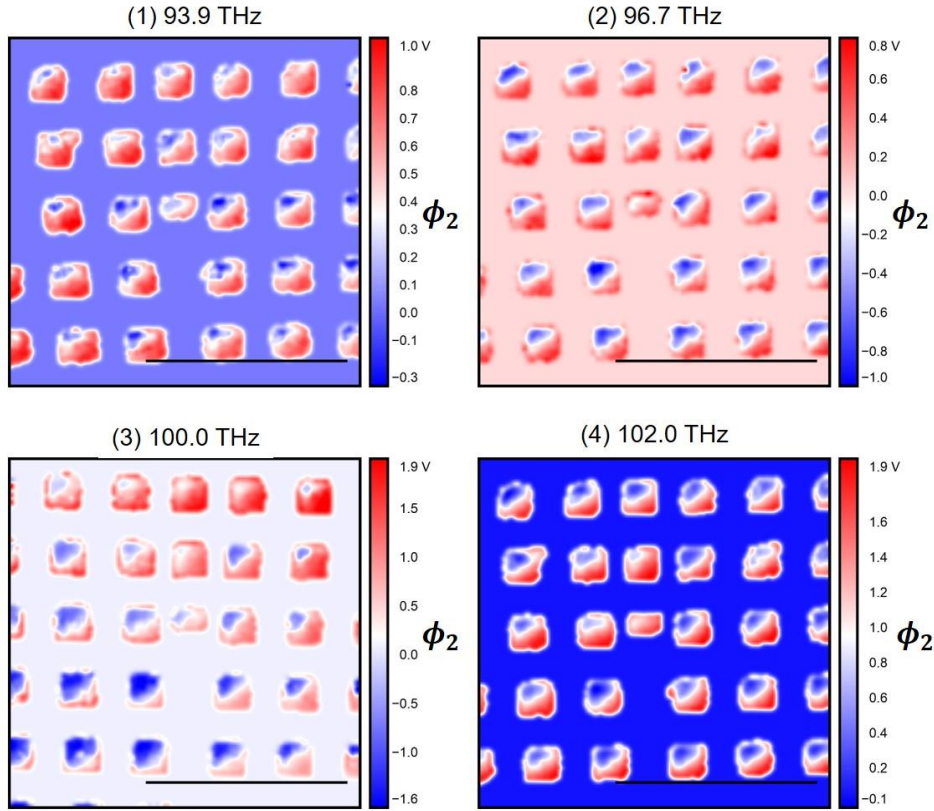

**Fig. S4.** s-SNOM second demodulation order phase images $\phi_2$ of the array depicted in Fig. 4 of the main text. The phase images were obtained in the same measurement as the amplitude images. The phase signal does not show a pronounced difference between the different frequencies. Scalebars are 4 μm.

In Fig. S5(a), we show the $E_z$ field distributions around the dislocation center for the localized mode ($f$ = 96.7 THz), which are plotted with different planes at z-direction (for heights of 650 nm to 800 nm). The height of the pillars is 673 nm, and the brightest position corresponds to the center single pillar as shown in the corresponding area in (b). All the planes show a strong field enhancement around this single pillar, which proves the possibility to detect the localized mode by moving the s-SNOM tip around the pillars.

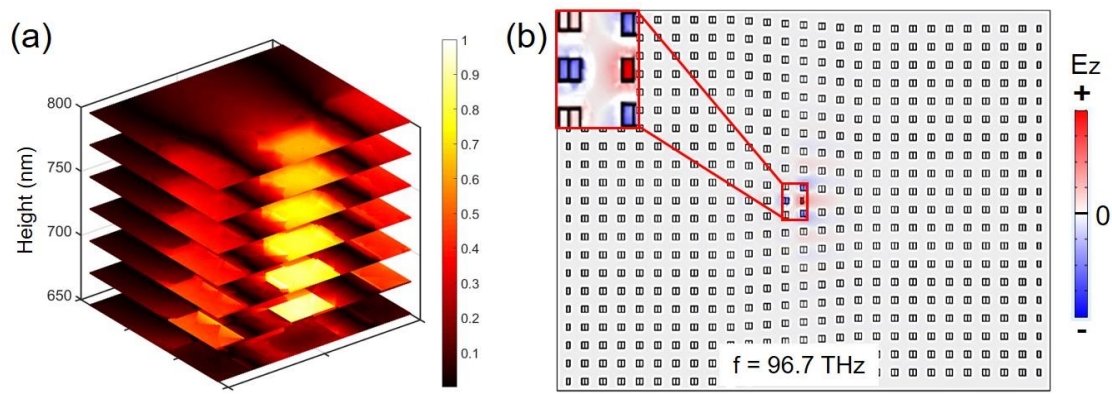

**Fig. S5.** (a) Normalized $|E_z|$ field distributions around the dislocation center (with the brightest position corresponding to the contribution from single pillar at the center for $f$ = 96.7 THz) plotted with different planes at z-direction (from 650 nm to 800 nm). (b) Simulated field distribution with the corresponding area plotted in (a) marked with a red box. All planes show a strong field enhancement around the dislocation center, which proves the possibility to detect the localized mode by scanning the s-SNOM tip around the pillars.

With the average $S_2$ s-SNOM amplitude signal on top of the pillar at the defect center (the green rectangle in Fig. S6(a)), we further normalize the data of each measured frequency use the value from the three pillars as indicated by white arrows (the three pillars two rows above the green rectangle). The highest observed amplitude corresponds to the localized frequency can be found as marked out with the black dotted line in Fig. S6(b). While the normalized amplitudes at 100 THz (inside the bandgap without localization) and other frequencies are much lower compared to the localized frequency.

To further add a quantitative evaluation of the measured light localization for different $N$ (including all the nontrivial TPhCs in the main text) by the same way. The ratio between the pillar at the defect center (the green rectangle) and the reference region (the three pillars two rows above in all the TPhCs) are shown in Fig. S6(c). The result shows a ratio increasement with an increase of defect center extension ($N$), which is consistent with the predication from simulation.

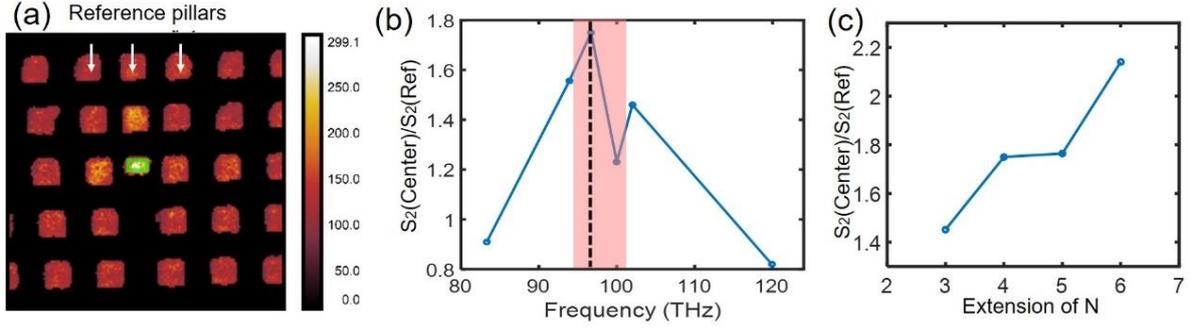

**Fig. S6.** (a) A $S_2$ s-SNOM amplitude image shows the data normalization arears. The average s-SNOM amplitude signal on top of the pillar at the defect center (the green rectangle) are normalized to the data from the reference (as indicated by the white arrows). (b) Normalized amplitudes are plotted against different measured frequencies. (c) The $S_2$ ratio between the pillar at the defect center and the reference region for different $N$.

Under background field excitation (nontrivial TPhC, $N = 4$), the nonmalized near-field ($Ez$) against the incident plane wave at the defect center (with the data from the top corner of the pillar) are plotted for different frequencies. A narrow resonant peak can be found whose frequency corresponds to the localized mode. The spectral information comfirms the topologically localized mode as a optical resonantor.

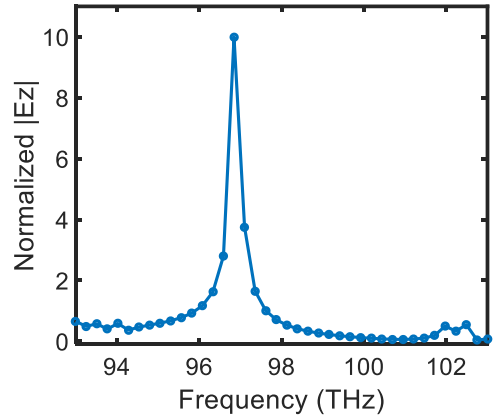

**Fig. S7.** Near-field ($Ez$) under plane wave excitation. A narrow resonant peak can be recongnized as for the in-bandgap localization.

### 4. Comparison between nontrivial and trivial TPhCs

We make the comparison between the nontrivial and trivial TPhCs for different $N$ as shown below. $N$ in both nontrivial and trivial TPhCs are based on the number of unit cell from the No. 10$^{th}$ one (see the column in green dotted box in Fig. S8(b)) at the center line. The corresponding trivial TPhC ($\theta = (0, 0)$) we choose here is because no dramatical change of unit cell at the center line for larger $N$. See Fig. 4E in main text, the two pillars at the center line is moved closely in the other trivial case ($\theta = (\pi/P, 0)$) even when $N = 4$, this could make the determanation of mode chellange.

According to principles of topology, only topological nontrivial modes can appear. The continuous wave nature of our system inevatibly introduce further complexities and allows the appearance of excessive trivial modes. However, we find that the topological mode's appearance follows the topological phase transition, while the trivial modes (if there are any) only experience minor frequency shifts. As is shown in Fig. S8, all the in-gap modes marked with red dots ($N$ = 3, 4, 5, 6 same with the make in main text) are nontrivial modes as they only exit in the nontrivial TPhCs. While the modes (marked out with red arrows in Fig. S8(b)) near the band edge in both spectra are trivial, as they exhibit similar frequency and field distribution for both trivial and nontrivial TPhCs.

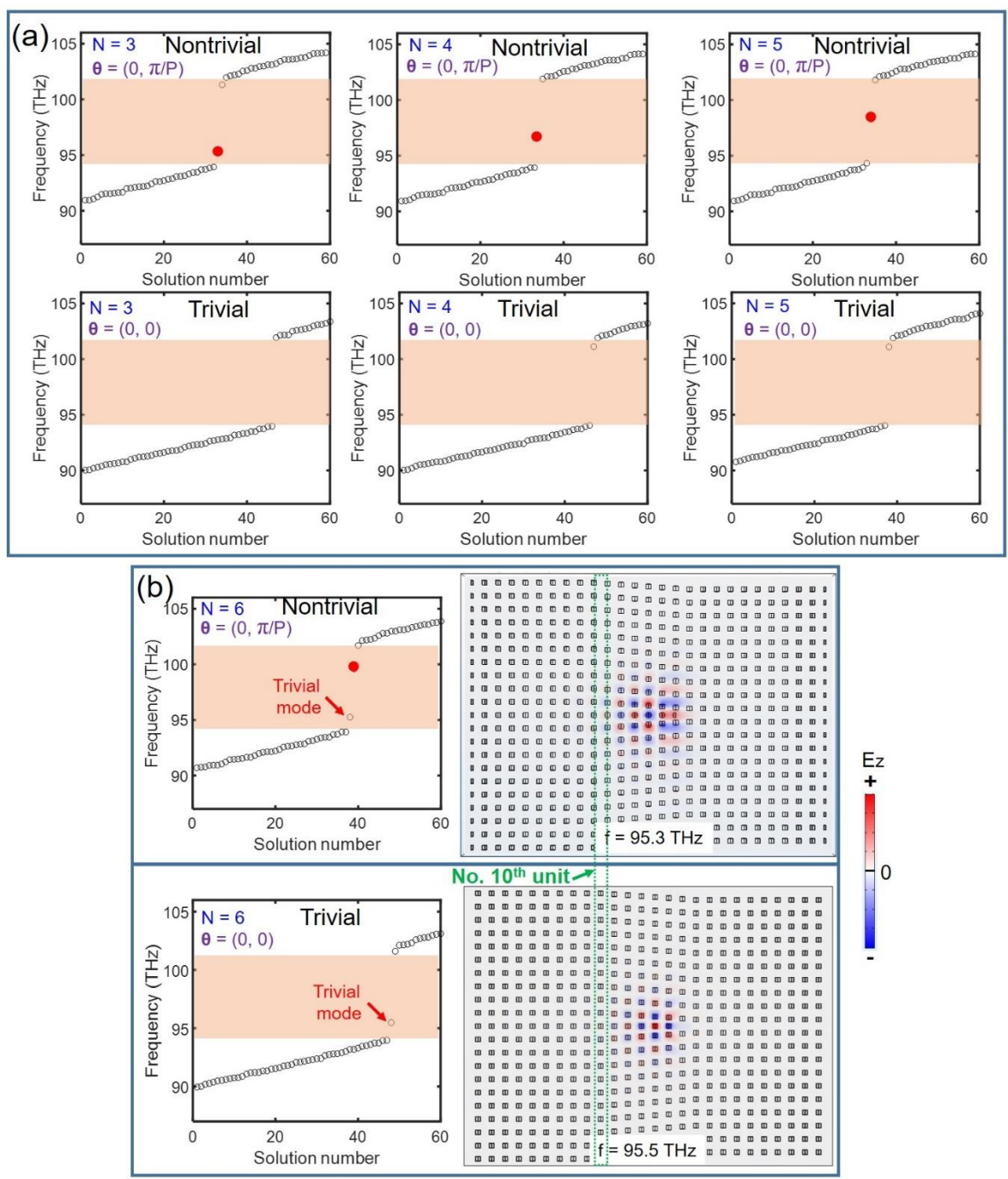

**Fig. S8.** (a) Nontrivial modes are marked with red dots for $N$ = 3, 4, 5. No in-bandgap trivial mode exists in the three corresponding trivial TPhCs in (a). (b) In-bandgap trivial modes for $N$ = 6 from both trivial and nontrivial TPhCs are indicated by red arrows. The $E_z$ field with similar distribution of these two trivial modes in (b) are shown on the right. The results show that these modes (we marked with red dots) only exist in nontrivial TPhCs are topologically protected. Both $\theta$ and $N$ are shown in each picture, and Burgers vectors for all the TPhCs are (0, $P$).

## 5. Robustness of the topologically protected mode

To investigate the robustness of the topological mode, we add random position perturbations to the pillars in both x and y directions and numerically simulate the nontrivial TPhC for $N = 4$ (corresponding to the TPhC in main text Fig. 4). With standard deviations (SD) of 50 nm (a) and 100 nm (b), it is found that the topologically protected localized mode remains with minor frequency shifts.

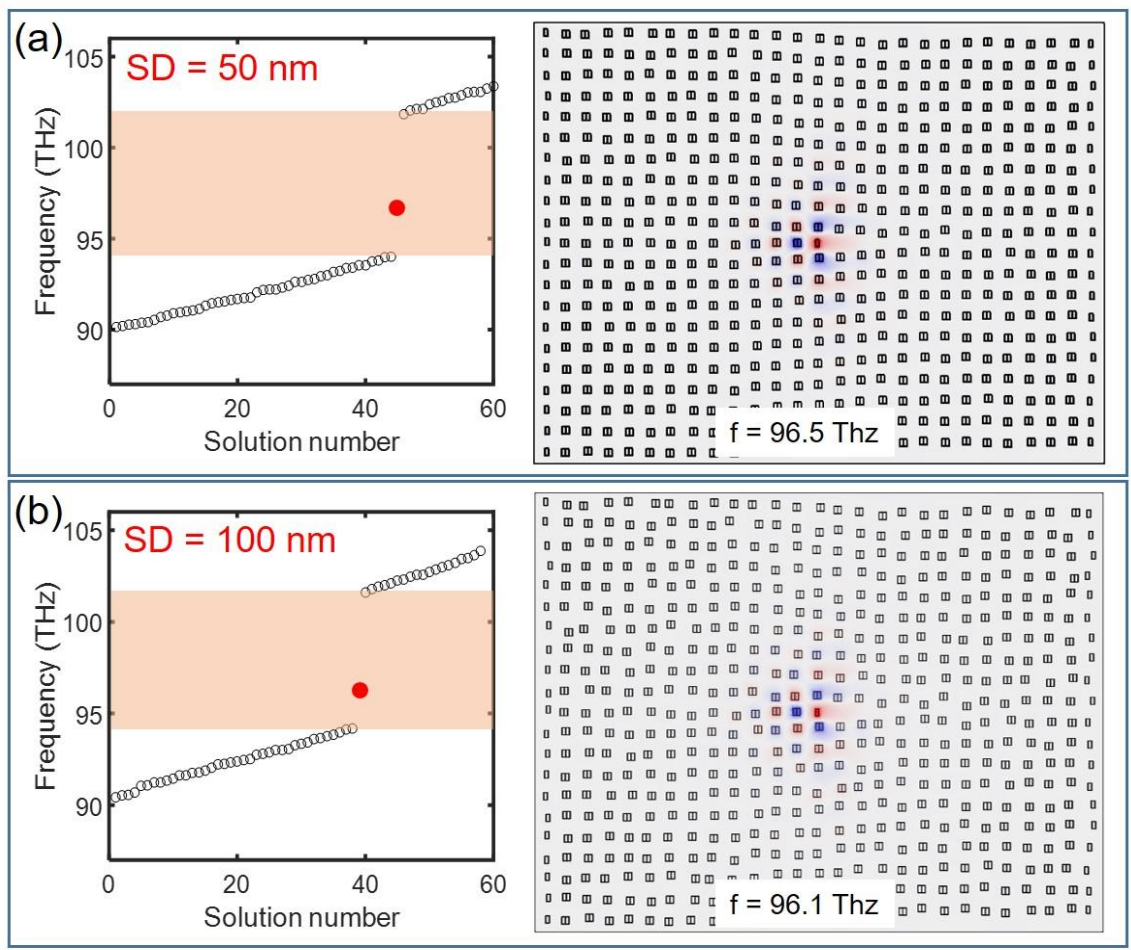

**Fig. S9.** Spectra of the nontrivial TPhC ($N = 4$) with the localized mode (red dot) in the bandgap after random position perturbations, (a) SD = 50 nm, (b) SD = 100 nm. $E_z$ field distributions corresponding to each in-gap localized mode in the spectrum are shown on the right.

## 6. Flow chart for the s-SNOM data processing procedure

With the measured result performed at 96.7 THz as an example, we show how to deal with the raw data of the s-SNOM amplitude measurements. First, the AFM signal is used to determine the top of the pillars by discriminating below a certain height. Then, the optical amplitude signal is obtained at a position where the height is above this threshold, and the amplitude value is set to zero at lateral positions below this threshold. This yields the amplitude signal on top of the pillars (Fig. S10(c)). To make the amplitude signals comparable between different frequencies the signal needs to be normalized. Thus, the signal on top of the middle pillar at the defect center (the green rectangle) is used for normalization and obtain the normalized optical amplitude image (Fig. S10(e)). Finally, a Gaussian filter is applied to the image to reduce the noise and the color bar is scaled from 0 to 1. The same procedure was applied to all the amplitude images in Fig. 4 and Fig. 5. Most of the background signal from the bottom is removed with this procedure.

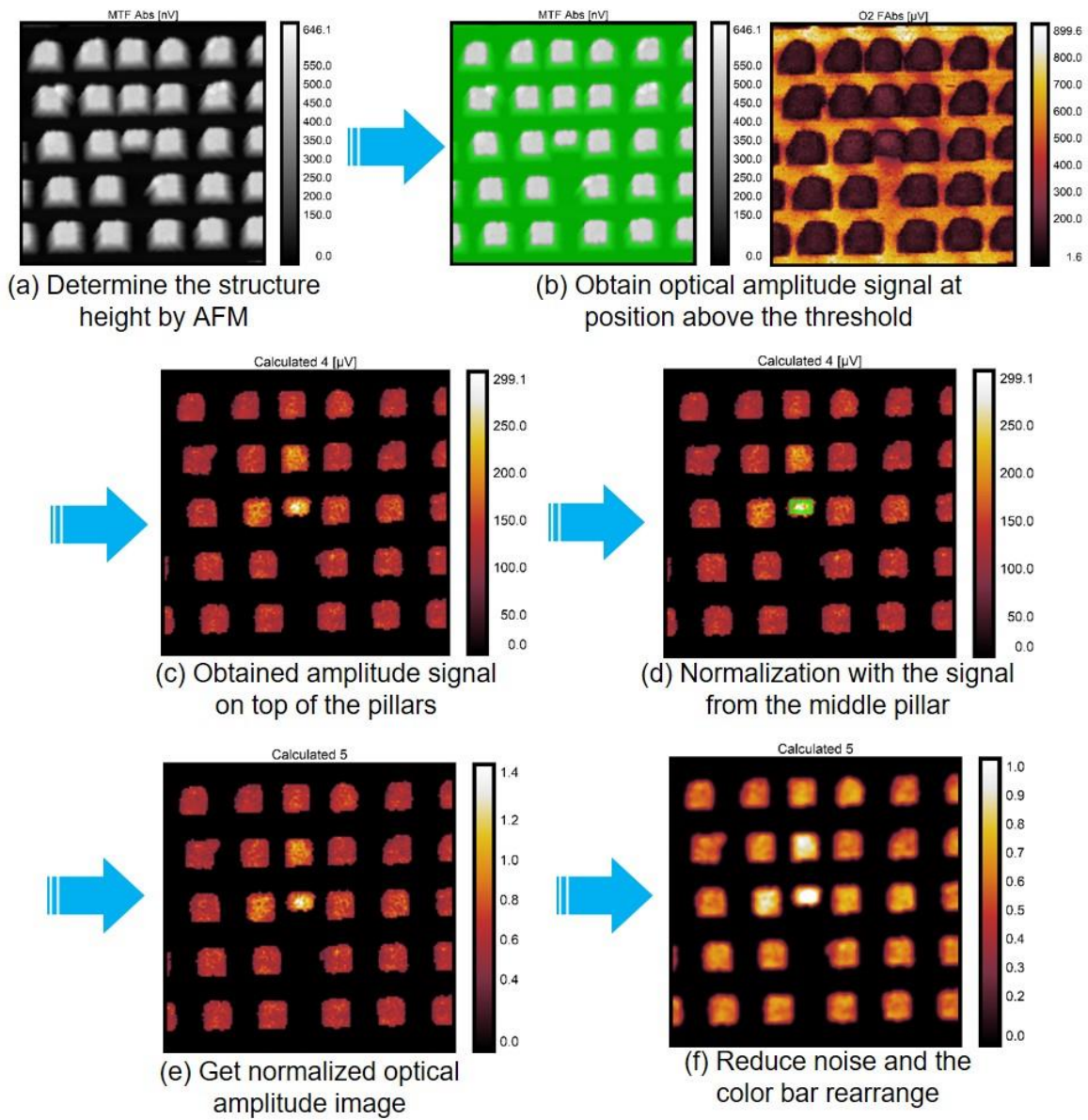

**Fig. S10.** Flow chart for the processing of the raw data from s-SNOM amplitude measurements ($S_2$). (a) The raw data of the s-SNOM's AFM. (b) Left: AFM image with a threshold mask to determine the top of the pillars. Right: $S_2$ optical amplitude image, where the very high optical signal between the pillars is because the interaction between the s-SNOM's tip and the Chromium at the bottom. (c) The obtained amplitude signal on top of the pillars. (d) The referencing region is indicated by the green box. (e) The normalized optical amplitude image. (f) Final normalized optical amplitude image with a Gaussian filter applied to reduce noise. The signal on the middle pillar in all images is around 1 after processing. Thus, the contrast from the pillar at the defect center to the surrounding pillars indicates the relative field strength on top of the pillars.

**Supplementary References**